\documentclass[twocolumn,prd,aps,nofootinbib,showpacs]{revtex4}
\usepackage{multirow}
\usepackage{graphicx}
\usepackage[dvips]{color}
\usepackage{amssymb,amsmath}

\begin{document}

\title{Lattice calculation of $\kappa$ meson}

\author{Ziwen Fu}

\email{fuzien@scu.edu.cn}

\affiliation{
Key Laboratory of Radiation Physics and Technology {\rm (Sichuan University)},
Ministry of Education; \\
Institute of Nuclear Science and Technology, Sichuan University,
Chengdu 610064, P. R. China.
}

\begin{abstract}
We study the $\kappa$ meson in $2+1$ flavor QCD with sufficiently light $u/d$ quarks.
Using numerical simulation we measure the point-to-point $\kappa$ correlators
in the ``Asqtad'' improved staggered fermion formulation.
We analyze those correlators
using the rooted staggered chiral perturbation theory (rS$\chi$PT),
particular attention is paid to the bubble contribution.
After chiral extrapolation, we obtain the physical $\kappa$ mass with $828\pm97$~MeV,
which is within the recent experimental value $800\sim900$MeV.
These numerical simulations are carried out with MILC $2+1$ flavor gauge configurations
at lattice spacing $a \approx 0.12$~fm.
\end{abstract}

\pacs{12.38.Gc,  11.15.Ha, 12.40.Yx, 14.40.Df}

\maketitle

\section{Introduction}
\label{sec_intro}
The so-called $\kappa$ meson ($J^P=0^+$) is a scalar meson with strangeness.
In 2010, Particle Data Group (PDG)~\cite{Nakamura:2010zzi} lists
the meson $K_0^*(800)$ or $\kappa$ with a very broad width ($550$ MeV).
A recent analysis \cite{Bugg:2005xx} gives its mass about $750^{30}_{50}$~MeV.
Moreover, the resonance of a scalar meson
is reported~\cite{Bugg:2005xx,Aitala:2002kr,Ablikim:2010kd,Ablikim:2005ni}
to exist in the $\pi K$ system with a mass of the $\kappa$ meson about 800 MeV.

Until now, four lattice simulations of the $\kappa$ mass have been reported.
Prelovsek {\it et al.} reported a crude estimation of
the $\kappa$ mass as $1.6$ GeV by extrapolating
the $a_0$ mass\emph{}~\cite{Prelovsek:2004jp}.
Mathur {\it et al.}~\cite{Mathur:2006bs} studied the $u\bar{s}$ scalar meson
in the quenched approximation, and
obtained the value of the $\kappa$ mass to be $1.41\pm0.12$~GeV
after removing the fitted $\pi\eta^{\prime}$ ghost.
With the dynamical $N_{f}=2$ sea quarks and a valence strange quark,
the UKQCD Collaboration~\cite{McNeile:2006nv}
suggested the $\kappa$ mass around $1000-1200$~MeV.
The SCALAR Collaboration~\cite{Kunihiro:2003yj,Kunihiro:2004ga}
reported full QCD simulations on the $\kappa$ meson
using the dynamical fermion for the light $u/d$ quark
and the valence approximation for the strange quark,
which shows that the $I=1/2$ scalar meson has a mass around $1.8$ GeV.
In Ref.~\cite{Wada:2007cp}, they performed a quenched QCD calculation
using the Wilson fermions with the plaquette gauge action,
and estimated the $\kappa$ mass to be about $1.7$ GeV.

In the presence of 2+1 flavors of Asqtad improved staggered dynamical sea quarks,
we obtained the $\kappa$ mass as low as $826 \pm 119$ MeV
in our previous study~\cite{fzw:2011cpc12},
however, here we neglected the taste-symmetry breaking
and used the crude linear extrapolation.
It is well known that, in the staggered fermion formulation of lattice QCD,
due to the taste-symmetry breaking there exist many multihadron states
with $J^P=0^+$ which can proliferate
between the source and sink of the $\kappa$ correlator.
Of special interest for us are the two-pseudoscalar states
(i.e., bubble contribution)~\cite{Prelovsek:2005qc,Prelovsek:2005rf}.
With sufficiently light $u$ quark, the $\kappa$ meson propagator
is dominated at large time distances by these two-meson states.
The bubble contributions are significantly affected
by the unphysical approximations which are often used
in lattice simulations~\cite{Prelovsek:2005qc,Prelovsek:2005rf}.
They are expected to disappear in the continuum limit.

In our previous work~\cite{Ph.D:2006fzw,Ph.D:2007fzw,fzw:2011cpl,fzw:2011cpc10},
we extended the analysis of Refs.~\cite{Prelovsek:2005qc,Prelovsek:2005rf},
examined the scalar meson correlators in lattice QCD with the inclusion
of the disconnected diagrams, and carried out a quantitative comparison of
the measured correlators and the predictions from rS$\chi$PT.
Despite the considerable complexity of the scalar meson channels with
dozens of spectral components, the rS$\chi$PT provides a strict framework
which permits the analysis of the scalar meson correlator
precisely in terms of only a small number of low-energy chiral couplings,
which we may determine through fits to the data.
In this work we extend the analyses of Refs.~\cite{Ph.D:2006fzw,Ph.D:2007fzw,fzw:2011cpl,fzw:2011cpc10}
to the $\kappa$ meson,
treat the $u$ quark as a valence approximation quark,
while the valence strange quark mass is fixed to its physical
value~\cite{Aubin:2004fs} considering that the $\kappa$ meson contains
a strange quark and a light $u$ quark,
we perform a series of numerical simulations
with MILC gauge configurations in the presence of $2+1$ flavors of
Asqtad improved staggered dynamical sea quarks,
generated by the MILC Collaboration~\cite{Bazavov:2009bb},
and chirally extrapolate the mass of the $\kappa$ meson
to the physical $\pi$ mass using the popular four parameter fit
with the inclusion of the chiral logarithms.

\section{Pseudoscalar meson taste multiplets}
\label{sec_taste_multiplets}
In Refs.~\cite{Ph.D:2006fzw,Ph.D:2007fzw},
we give a brief review of the rooted staggered chiral perturbation theory
with particular focus on the tree-level pseudoscalar mass spectrum,
and achieve the rooted version of the theory
through the replicated theory~\cite{Aubin:2003rg}.

The tree-level masses of the pseudoscalar mesons are~\cite{Ph.D:2007fzw,Aubin:2003mg}
\begin{equation}
M_{x,y,b}^2 = \mu(m_x + m_y ) + a^2\Delta_b,
\label{tree_splitting}
\end{equation}
where $x, y$ are two quark flavor contents which make up,
$b=1,...,16$ are the taste,
$\displaystyle \mu= m_\pi^2/2m_q$ is the low-energy chiral coupling constant
of the point scalar current to the pseudoscalar field,
and the term of $a^2\Delta_b$ comes from the taste symmetry breaking.
The $m_x$ and $m_y$ are the two valence quark masses in the
pseudoscalar meson, and $m_x$ is the light valence $u/d$ quark mass by convention.

In this work we investigate degenerate $u$ and $d$ quarks,
treat $u$ quark as a valence approximation quark,
while the valence strange quark mass is fixed to its physical value $m_s$,
thus it will be convenient to introduce the notations
\begin{eqnarray}
  M^2_{Ub}  \hspace{-0.1cm} &\equiv& \hspace{-0.1cm} M_{\pi_b} =
  2\mu m_x  + a^2 \Delta_b  \nonumber \\
  M^2_{Sb}  \hspace{-0.1cm} &\equiv& \hspace{-0.1cm} M_{ss,b} =
  2\mu m_s  + a^2 \Delta_b  \\
  M^2_{Kb}  \hspace{-0.1cm} &\equiv& \hspace{-0.1cm} M_{K_b} =
  \mu (m_x  + m_s)  + a^2 \Delta_b, \nonumber
\end{eqnarray}
where $M_U$ is the mass of the Goldstone pion with two light valence quark masses,
$M_K$ is the mass of the Goldstone kaon with one valence quark equal
to the light valence quark and one at its physical mass $m_s$, and
$M_S$ is the mass of a fictitious meson $s\bar{s}$ in a flavor nonsinglet
state~\cite{Aubin:2004wf} with two valence quarks at physical mass $m_s$.

The isosinglet states ($\eta$ and $\eta^\prime$) are modified both by
the taste-singlet anomaly and by the two-trace (quark-line hairpin)
taste-vector and taste-axial-vector operators \cite{Bernard:2001av,Aubin:2004wf}.
When the anomaly parameter $m_0$ is large, we obtain the usual result
\begin{equation}
M_{\eta, I}^2 = \frac{1}{3}M_{UI}^2 + \frac{2}{3}M_{SI}^2, \hspace{0.5cm}
M_{\eta^\prime, I} =  {\cal O}(m_0^2).
\end{equation}
In the taste-axial-vector sector we have
\begin{eqnarray}
M^2_{\eta A}\hspace{-0.1cm}&=&\hspace{-0.1cm}\frac{1}{2}[M^2_{UA}+M^2_{SA}+\frac{3}{4}\delta_A-Z_A] \nonumber \\
M^2_{\eta^\prime A}\hspace{-0.1cm}&=&\hspace{-0.1cm}\frac{1}{2}[M^2_{UA}+M^2_{SA}+\frac{3}{4}\delta_A+Z_A] \\
Z^2_A\hspace{-0.1cm}&=&\hspace{-0.1cm}(M^2_{SA}-M^2_{UA})^2-\frac{\delta_A}{2} (M^2_{SA}-M^2_{UA})+\frac{9}{16}\delta^2_A, \nonumber
\label{eq_etaAV}
\end{eqnarray}
and likewise for $V \to A$,
where $\delta_V = a^2\delta_V^{\prime}$ is the hairpin coupling
of a pair of taste-vector mesons,
and $\delta_A = a^2\delta_A^{\prime}$ is the hairpin coupling of
a pair of taste-axial mesons
($\delta_V^{\prime}$ and $\delta_A^{\prime}$ are tree-level (LO)
taste-violating hairpin parameters~\cite{Aubin:2004fs}).

\begin{table}
\caption{ \label{tab_L0}
The mass spectrum of the pseudoscalar meson for MILC medium-coarse
($a = 0.12$ fm) lattice ensemble
with $\beta = 6.76$, $am_{ud}^{\prime} = 0.005$, $am_s^{\prime} = 0.05$.
}
\begin{ruledtabular}
\begin{tabular}{cccccc}
$am_x$ & taste($B$) & $a\pi_B$ & $aK_B$ & $a\eta_B$ & $a\eta^\prime_B$ \\
\hline
\multirow{5}*{$0.005$}
&P & $0.1598$ & $0.3106$ & $0.1598$ & $0.4087$ \\
&A & $0.2342$ & $0.3546$ & $0.1831$ & $0.4332$ \\
&T & $0.2688$ & $0.3784$ & $0.2688$ & $0.4624$ \\
&V & $0.2971$ & $0.3990$ & $0.2832$ & $0.4755$ \\
&I & $0.3198$ & $0.4162$ & $0.4434$ & $-$ \\
\hline
\multirow{5}*{$0.010$}
&P & $0.2233$ & $0.3291$  &$0.2233$ & $0.4087$ \\
&A & $0.2814$ & $0.3710$  &$0.2400$ & $0.4335$ \\
&T & $0.3108$ & $0.3938$  &$0.3108$ & $0.4624$ \\
&V & $0.3356$ & $0.4136$  &$0.3233$ & $0.4755$ \\
&I & $0.3558$ & $0.4302$  &$0.4525$ & $-$ \\
\hline
\multirow{5}*{$0.015$}
&P & $0.2718$ & $0.3468$  & $0.2718$ & $0.4087$ \\
&A & $0.3212$ & $0.3867$  & $0.2851$ & $0.4339$ \\
&T & $0.3473$ & $0.4087$  & $0.3473$ & $0.4624$ \\
&V & $0.3696$ & $0.4278$  & $0.3584$ & $0.4756$ \\
&I & $0.3881$ & $0.4439$  & $0.4612$ & $-$      \\
\hline
\multirow{5}*{$0.020$}
&P & $0.3127$ & $0.3637$  & $0.3127$ & $0.4087$ \\
&A & $0.3565$ & $0.4020$  & $0.3235$ & $0.4345$ \\
&T & $0.3802$ & $0.4231$  & $0.3802$ & $0.4624$ \\
&V & $0.4007$ & $0.4416$  & $0.3902$ & $0.4757$ \\
&I & $0.4178$ & $0.4572$  & $0.4698$ & $-$      \\
\hline
\multirow{5}*{$0.025$}
&P & $0.3489$ & $0.3798$  & $0.3489$ & $0.4087$ \\
&A & $0.3886$ & $0.4166$  & $0.3572$ & $0.4356$ \\
&T & $0.4105$ & $0.4370$  & $0.4105$ & $0.4624$ \\
&V & $0.4295$ & $0.4550$  & $0.4195$ & $0.4759$ \\
&I & $0.4455$ & $0.4701$  & $0.4782$ & $-$\\
\end{tabular}
\end{ruledtabular}
\end{table}

In the taste-pseudoscalar and taste-tensor sectors,
in which there is no mixing of the isosinglet states,
the masses of the $\eta_b$ and $\eta'_b$ by definition are
\begin{equation}
M_{\eta, b}^2 = M_{Ub}^2 ; \qquad  M_{\eta', b}^2 = M_{Sb}^2 .
\end{equation}
In Table~\ref{tab_L0}, we list the masses of the resulting taste multiplets
in lattice units for our chosen lattice ensemble
with the taste-breaking parameters $\delta_A$ and $\delta_V$
determined in Refs.~\cite{Aubin:2004fs,Aubin:2004wf}.
For Goldstone multiplet (taste $P$), we measured
their corresponding correlators and fitted them
with a single-exponential~\cite{Aubin:2004fs}.
Then, using the taste splittings in Refs.~\cite{Aubin:2004fs,Aubin:2004wf},
we calculated the masses of other non-Goldstone taste multiplets.
We do not need the $\eta^\prime_I$ masses in this work,
hence, we do not list these values in Table~\ref{tab_L0}.

We should remind the readers that the data in Table~\ref{tab_L0}
are obtained with the valence strange quark mass fixed to its physical value.
Here and below, we adopt the notation in Ref.~\cite{Aubin:2004fs},
the primes on masses indicate that they are the dynamical
quark masses used in the lattice simulations, not the physical
masses $m_u$, $m_d$ and $m_s$.

\section{The $\kappa$ correlator from S$\chi$PT}
\label{sec:correlators}
In Refs.~\cite{Ph.D:2006fzw,Ph.D:2007fzw,fzw:2011cpl,fzw:2011cpc10},
using the language of the replica trick~\cite{Damgaard:2000gh,Aubin:2003mg} and
through matching the point-to-point scalar correlators in chiral low energy
effective theory and staggered fermion QCD,
we rederive the ``bubble'' contribution to
the $a_0$ channel of Ref.~\cite{Prelovsek:2005rf},
and extend the result to the $\sigma$ channel.
Here we further extend these results to the $\kappa$ channel.

\subsection{ Non-singlet $\kappa$ correlator }
To simulate the correct number of quark species,
we use the fourth-root trick, which automatically performs the transition
from four tastes to one taste per flavor for staggered fermion at all orders.
We employ an interpolation operator with isospin $I=1/2$ and $J^{P}=0^{+}$
at the source and sink,
\begin{eqnarray}
{\cal O}(x)  \equiv
\frac{1}{\sqrt{n_r}} \sum_{a, g}\bar s^a_g( x ) u^a_g(x) ,
\end{eqnarray}
where $g$ is the indices of the taste replica,
      $n_r$ is the number of the taste replicas, $a$ is the color indices,
      and we omit the Dirac-Spinor index.
The time slice correlator $C(t)$ for the $\kappa$ meson can be evaluated by
\begin{eqnarray}
C(t)  \hspace{-0.3cm} &=& \hspace{-0.3cm}
\frac{1}{n_r}
\sum_{ {{\textit{\textbf{ x} }}}, a, b} \sum_{g, g'}
\left\langle \bar s^{b}_{g'}(\hspace{-0.18cm}{{\textit{\textbf{ x} }}}, t)
u^{b}_{g'}(\hspace{-0.18cm}{{\textit{\textbf{ x} }}}, t) \
\bar u^{a}_{g }({\bf 0}, 0) s^{a}_{g }({\bf 0}, 0) \right\rangle , \nonumber
\label{EQ.kappa}
\end{eqnarray}
where ${\bf 0, \hspace{-0.15cm}{{\textit{\textbf{ x} }}}}$ are the spatial points of
the $\kappa$ state at source, sink, respectively.
After performing Wick contractions of fermion fields,
and summing over the taste index~\cite{fzw:2011cpc12},
for the light $u$ quark Dirac operator $M_u$ and the $s$ quark Dirac operator $M_s$, we obtain
\begin{equation}
\label{CCEQ_kappa}
C(t) = \sum_{ {\textit{\textbf{ x} }} }(-)^x
\left\langle \mbox{Tr}
[M^{-1}_u(\hspace{-0.15cm}{{\textit{\textbf{ x} }}},t;0,0)
M^{-1^\dag}_s( \hspace{-0.15cm}{{\textit{\textbf{ x} }}},t;0,0)] \right\rangle ,
\end{equation}
where $\mbox{Tr}$ is the trace over the color index,
and $x=(\hspace{-0.18cm}{{\textit{\textbf{ x} }}},t)$ is the lattice position.

The mass of the $\kappa$ meson can be reliably
determined on the lattice simulation.
However, there exist many multihadron states with $J^P=0^+$
which can propagate between the source and the sink.
Of special interest for us is the intermediate state
with two pseudoscalars $P_1 P_2$ which we refer to as
the bubble contribution ($B$)~\cite{Prelovsek:2005rf}.
If the masses of $P_1$ and $P_2$ are small,
the bubble term gives a considerable contribution
to the $\kappa$ correlator, and it should be included
in the fit of the lattice correlator in Eq.~(\ref{CCEQ_kappa}), namely,
\begin{equation}
\label{Ctot_kappa}
C(t)=Ae^{-m_{\kappa}t}+B(t),
\end{equation}
where we omit the unimportant contributions from the excited states,
the oscillating terms corresponding to a particle with opposite parity,
and other high-order multihadron intermediate states.

\subsection{Coupling of a scalar current to pseudoscalar}
Before we embark on the bubble contribution,
here we first derive the coupling of a point scalar current
$\bar s_r(x)u_r(x)$ to a pair of the pseudoscalar fields at the lowest
energy order of the staggered chiral perturbation theory (S$\chi$PT),
where the subscript $r$ in the expression $u_r(x)$ is
the index of the taste replica for a given quark flavor $u$.
The effective scalar current can be determined from the dependence
of the lattice $QCD$ Lagrangian and the staggered chiral Lagrangian on
the spurion field ${\cal M}$, where ${\cal M}$ is the
staggered quark mass matrix. For $n$  Kogut-Susskind (KS) flavors,
${\cal M}$ is a $4n n_r \times 4n n_r $ matrix.
\begin{eqnarray}
{\cal M} \hspace{-0.1cm}= \hspace{-0.1cm}\left( \begin{array}{cccc}
m_{uu} I\otimes I_R&m_{ud} I\otimes I_R&m_{us} I \otimes I_R& \cdots \\*
m_{du} I\otimes I_R&m_{dd} I\otimes I_R&m_{ds} I \otimes I_R& \cdots \\*
m_{su} I\otimes I_R&m_{sd} I\otimes I_R&m_{ss} I \otimes I_R& \cdots \\*
        \vdots & \vdots & \vdots & \ddots \end{array} \right) ,
\end{eqnarray}
where $I$ is a $4 \times 4$ unit matrix, and
$I_R$ is the $ n_r \times n_r $ unit replica matrix.
In short, ${\cal M} =  {\bf m } \otimes I \otimes I_R $, and
\begin{eqnarray}
{\bf m} = \left( \begin{array}{cccc}
	 m_{uu}   & m_{ud}    & m_{us}     & \cdots \\*
	 m_{du}   & m_{dd}    & m_{ds}     & \cdots \\*
	 m_{su}   & m_{sd}    & m_{ss}     & \cdots \\*
      \vdots & \vdots & \vdots & \ddots \end{array} \right)
\end{eqnarray}
is the $n\times n$ quark mass matrix.
Since the staggered chiral lagrangian (${\cal L}_{S \chi PT}$)
is an effective equivalent Lagrangian for ${\cal L}_{\rm QCD}$ in
low energy limit, the effective current $\bar s_r(x)u_r(x)$ is obtained from
\begin{equation}
\bar{s}_r(x)u_r(x) =
-\frac{ \partial{\cal L}_{S \chi PT}}{\partial{\cal M}_{s_r u_r}(x)} ,
\end{equation}
where
\begin{equation}
{\cal L}_{S \chi PT} = \frac{1}{8} f_{\pi}^2 {\rm Tr}
[\partial^\mu \Sigma \partial_\mu \Sigma^\dag] - \frac{1}{4}
 \mu f_{\pi}^2{\rm Tr}[{\cal M}^\dag \Sigma+\Sigma^\dag{\cal M}] \nonumber\
\end{equation}
is the staggered chiral Lagrangian~\cite{Aubin:2003mg}.
We omit the high order terms and the terms that are independent of ${\cal M}$,
$f_{\pi}$ is the tree-level pion decay constant~\cite{Aubin:2003mg}, $\mu$
is the constant with the dimension of the mass~\cite{Bardeen:2001jm},
and $\displaystyle \Sigma=\exp \left( \frac{2i\Phi}{f_{\pi}} \right)$~\cite{Aubin:2003mg}.
The field $ \displaystyle \Phi=\sum_{b=1}^{16} \frac{1}{2}T^b \otimes \phi^b$
is described in terms of the mass eigenstate field $\phi^b$~\cite{Aubin:2003mg},
where $\phi^b$ is a $3 \times 3$ pseudoscalar matrix with flavor
components $\phi^b_{ f_r f'_{r'} }$ with flavor $f,f'$, the index of
the taste replica $r, r'$, and taste $b$ which is given by generators
$T^b=\{\xi_5,i\xi_5\xi_\mu,i\xi_\mu\xi_\nu,\xi_\mu,\xi_I\}$~\cite{Aubin:2003mg}.
Hence, the field $\Phi$ is $ 4 n n_r \times 4n n_r $ pseudoscalar matrix
in S$\chi$PT~\cite{Aubin:2003mg}, and the subscripts $u,s$ denote
its valance flavor component.
Therefore, ${\Sigma}$ is also a $4n n_r \times 4n n_r$ matrix.
The {\rm Tr} is the full $4n n_r \times 4n n_r$ trace.
Then, the effective current is~\cite{Bardeen:2001jm}
\begin{equation}
\bar s_r(x) u_r(x) =
\mu {\rm Tr_t } [\Phi(x)^2]_{s_r u_r} ,
\end{equation}
where the notation ${\rm Tr_t}$ stands for the trace over taste,
and $\mu$ is the low-energy chiral coupling of the point scalar current
to the pseudoscalar field $\Phi(x)$.

\subsection{Bubble contribution to the $\kappa$ correlator}
In this subsection we compute the bubble contribution to $\kappa$ correlator
in Eq.~(\ref{CCEQ_kappa}) from two intermediate states.
From the above discussion, the point scalar current
can be described in terms of the pseudoscalar field
$\Phi$ by using S$\chi$PT~\cite{Bardeen:2001jm}\cite{Prelovsek:2005rf},
\begin{eqnarray}
\label{current::appC}
\bar  s_r(x) u_r(x)  &=&  \mu {\rm Tr_t}[\Phi(x)^2]_{s_r u_r} , \nonumber\\
\bar  u_r(x) s_r(x)  &=&  \mu {\rm Tr_t}[\Phi(x)^2]_{u_r s_r} .
\end{eqnarray}

For concreteness, the bubble contribution to the $\kappa$ correlator
in a theory with $n_r$ tastes per flavor and  three flavors
of KS dynamical sea quarks is~\cite{Aubin:2003mg}
\begin{eqnarray}
B^{S\chi PT}_{\kappa}(x) &\hspace{-0.3cm}=\hspace{-0.3cm}&
\frac{\mu^2}{n_r}  \sum_{r,r'=1}^{n_r}
\Bigg\{
\left\langle \bar s_r(x) u_r(x) \ \bar u_{r^\prime}(0) s_{r^\prime}(0)
\right\rangle_{\rm Bubble}  \Biggr\} \nonumber \\
 &\hspace{-0.3cm}=\hspace{-0.3cm}&
\frac{\mu^2}{n_r} \sum_{r,r^\prime=1}^{n_r}  \Bigg\{
\left\langle{\rm Tr_t}[\Phi^2]_{s_r u_r}
            {\rm Tr_t}[\Phi^2]_{u_{r^\prime} s_{r^\prime}}
\right\rangle \Biggr\} , \nonumber
\end{eqnarray}
where the subscript $\kappa$ specifies the bubble contribution for $\kappa$ meson.
If we consider this identity
\begin{equation}
{\rm Tr_t} \left( T^{a} T^{b} \right) = 4  \delta_{ab},\nonumber
\end{equation}
we arrive at
\begin{eqnarray}
B^{S \chi PT}_{\kappa}(x) \hspace{-0.1cm} &=& \hspace{-0.1cm}
\frac{\mu^2}{n_r}
\sum_{a=1}^{16}
\sum_{b=1}^{16}
\sum_{i=1}^{N_f}
\sum_{j=1}^{N_f}
\sum_{r,r'=1}^{n_r}
\sum_{t,t'=1}^{n_r}
\nonumber \\
 \hspace{-0.2cm}  && \hspace{-0.2cm}
\langle
\phi^a_{  s_r    i_t   }(x) \phi^a_{i_t      u_r}(x) \
\phi^{b}_{u_{r^\prime} j_{t^\prime}}(0) \phi^{b}_{j_{t^\prime} s_{r^\prime} }(0)
\rangle,\nonumber
\end{eqnarray}
where $a,b,a^\prime,b^\prime$ are the taste indices, $i,j$ are the flavor indices,
and   $r,t,r^\prime,t^\prime$ are the indices of the taste replica.
The Wick contractions result in the products of two propagators
for the pseudoscalar fields.
After summing over the index of the taste replica,
and considering that $u,d$ quarks are degenerate,
the bubble contribution can be expressed in terms of the
pseudoscalar propagators  $\langle \phi^b \ \phi^b\rangle$.
\begin{widetext}
\begin{eqnarray}
B^{S \chi PT}_{\kappa}(x) \hspace{-0.1cm} &=& \hspace{-0.1cm}
n_r \mu^2 \sum_{b=1}^{16} \Biggl\{
 \langle \phi^b_{sd}(x)  \phi^b_{us}(0) \rangle_{\rm conn}
 \langle \phi^b_{ss}(x)  \phi^b_{ss}(0) \rangle_{\rm conn} +
 \langle \phi^b_{su}(x)  \phi^b_{us}(0) \rangle_{\rm conn}
 \langle \phi^b_{uu}(x)  \phi^b_{uu}(0) \rangle_{\rm conn} + \nonumber\\
  && \hspace{1.5cm}
 \langle \phi^b_{su}(x)  \phi^b_{us}(0) \rangle_{\rm conn}
 \langle \phi^b_{ss}(x)  \phi^b_{ss}(0) \rangle_{\rm conn}
\Biggr\} +  \nonumber\\
  && \mu^2 \sum_{b=I,V,A}
\Biggl\{
   \langle \phi^b_{su}(x)  \phi^b_{us}(0) \rangle_{\rm conn}
   \langle \phi^b_{uu}(x)  \phi^b_{uu}(0) \rangle_{\rm disc} +
   \langle \phi^b_{su}(x)  \phi^b_{us}(0) \rangle_{\rm conn}
   \langle \phi^b_{ss}(x)  \phi^b_{ss}(0) \rangle_{\rm disc} + \nonumber\\
  && \hspace{1.5cm} 2\langle \phi^b_{su}(x)  \phi^b_{us}(0) \rangle_{\rm conn}
   \langle \phi^b_{uu}(x)  \phi^b_{ss}(0) \rangle_{\rm disc}
 \Biggr\} .
\label{schpt_0:kappa:appC}
\end{eqnarray}
\end{widetext}
The subscript ${\rm conn}$ and subscript ${\rm disc}$ stand for
the connected contribution and the disconnected contribution, respectively.
The propagators $\langle \phi^b \phi^b\rangle$  for the pseudoscalar field
$\phi^b$ for various tastes $b$ are intensively studied in Ref.~\cite{Aubin:2003mg}.
The propagators for all tastes ($I,V,A,T,P$) have the connected contributions,
while only tastes $I,V$ and $A$ have the disconnected contributions~\cite{Aubin:2003mg}.
\begin{equation}
\langle \phi^b_{uu}(x)  \phi^b_{uu}(0) \rangle_{\rm disc}^V  =
 \frac{ -\delta_V(k^2 \hspace{-0.05cm}+\hspace{-0.05cm} M_{S_V}^2) }{
(k^2 \hspace{-0.05cm}+\hspace{-0.05cm} M_{U_V}^2)
(k^2 \hspace{-0.05cm}+\hspace{-0.05cm} M_{\eta_V}^2)
(k^2 \hspace{-0.05cm}+\hspace{-0.05cm} M_{\eta^\prime V}^2)}, \nonumber
\end{equation}
\begin{equation}
\langle \phi^b_{ss}(x)  \phi^b_{ss}(0) \rangle_{\rm disc}^V  =
\frac{ -\delta_V(k^2 \hspace{-0.05cm}+\hspace{-0.05cm} M_{U_V}^2) }{
(k^2 \hspace{-0.05cm}+\hspace{-0.05cm} M_{S_V}^2)
(k^2 \hspace{-0.05cm}+\hspace{-0.05cm} M_{\eta_V}^2)
(k^2 \hspace{-0.05cm}+\hspace{-0.05cm} M_{\eta^\prime V}^2)} ,\nonumber
\end{equation}
\begin{equation}
\langle \phi^b_{uu}(x)  \phi^b_{ss}(0) \rangle_{\rm disc}^V =
- \frac{\delta_V}{(k^2 + M_{\eta_V}^2)(k^2 + M_{\eta^\prime V}^2)},\nonumber
\end{equation}
\begin{equation}
\langle \phi^b_{uu}(x) \ \phi^b_{uu}(0) \rangle_{\rm disc}^I = -\frac{4}{3}
\frac{ k^2 + M_{S_I}^2 }{(k^2 + M_{U_I}^2)(k^2 + M_{\eta I}^2)}  ,\nonumber
\end{equation}
\begin{equation}
\langle \phi^b_{ss}(x) \ \phi^b_{ss}(0) \rangle_{\rm disc}^I =  -\frac{4}{3}
\frac{ k^2 + M_{U_I}^2 }{(k^2 + M_{S_I}^2)(k^2 + M_{\eta I}^2)} ,\nonumber
\end{equation}
\begin{equation}
\langle \phi^b_{uu}(x) \ \phi^b_{ss}(0) \rangle_{\rm disc}^I =  -\frac{4}{3}
\frac{1}{k^2 + M_{\eta I}^2} . \nonumber
\end{equation}
And likewise for the axial taste ($A$), we just require $V \to A$.
By plugging in the above mesonic propagators
(namely, carrying out the Wick contractions)
and switching to momentum space,
Eq.~(\ref{schpt_0:kappa:appC}) can be rewritten as
\begin{widetext}
\begin{eqnarray}
\label{schpt_kappa}
B^{S \chi PT}_{\kappa}(p) \hspace{-0.2cm} &=& \hspace{-0.2cm}
\mu^2 \sum_k  \Biggl\{
n_r \sum_{b=1}^{16}
\Biggl[2\frac{1}{(k + p)^2 + M_{K_b}^2} \frac{1}{k^2 + M_{U_b}^2}+
        \frac{1}{(k + p)^2 + M_{K_b}^2} \frac{1}{k^2 + M_{S_b}^2}
\Biggr]  \nonumber\\
&&\hspace{-1.2cm}
-2 \frac{1}{(k + p)^2 + M_{K_I}^2}\frac{1}{k^2 + M_{U_I}^2 }
+\frac{2}{3}\frac{1}{(k + p)^2 + M_{K_I}^2}\frac{1}{k^2 + M_{\eta I}^2}
-4\frac{1}{(k + p)^2 + M_{K_I}^2}\frac{1}{k^2 + M_{S_I}^2}\nonumber\\
&&\hspace{-1.2cm}-
\frac{ \delta_V}{ (k + p)^2 + M_{K_V}^2 }
\frac{ k^2 + M_{S_V}^2 }
{(k^2 + M_{M_V}^2)(k^2 + M_{\eta_V}^2)(k^2 + M_{\eta^\prime V}^2)}
-
\frac{\delta_A}{ (k + p)^2 + M_{K_A}^2 }
\frac{k^2+M_{S_A}^2}
{(k^2 + M_{U_A}^2)(k^2 + M_{\eta_A}^2)(k^2 + M_{\eta^{\prime} A}^2)}\nonumber\\
&&\hspace{-1.2cm}-
\frac{\delta_V}{ (k + p)^2 + M_{K_V}^2 }
\frac{ k^2 + M_{U_V}^2 }
{(k^2 + M_{S_V}^2)(k^2 + M_{\eta_V}^2)(k^2 + M_{\eta^\prime V}^2)}
-
\frac{\delta_A}{ (k + p)^2 + M_{K_A}^2 }
\frac{k^2 + M_{U_A}^2}
{(k^2 + M_{S_A}^2)(k^2 + M_{\eta_A}^2)(k^2 + M_{\eta^{\prime} A}^2)}\nonumber\\
&&\hspace{-1.2cm}-
2\frac{ \delta_V}{ (k + p)^2 + M_{K_V}^2 }
\frac{1}{(k^2 + M_{\eta_V}^2)(k^2 +  M_{\eta^\prime V}^2)}
-
2\frac{\delta_A}{ (k + p)^2 + M_{K_A}^2 }
\frac{1}{(k^2 + M_{\eta_A}^2)(k^2 + M_{\eta^{\prime} A}^2)}
\Biggr\} ,
\end{eqnarray}
\end{widetext}

And likewise for the axial taste ($A$), we just require $V \to A$.
In Table~\ref{tab_L0}, we list all these pseudoscalar masses
needed for the current study.

In the continuum limit, taste-symmetry is restored, namely,
$\delta_V = a^2\delta_V^{\prime} \to 0 $, $\delta_A = a^2\delta_A^{\prime} \to 0$,
then Eq.~(\ref{schpt_kappa}) reduces to
\begin{eqnarray}
B_{\kappa}(p) \hspace{-0.2cm} &=& \hspace{-0.2cm}
\mu^2 \sum_k
\Biggl\{
   2n_r \sum_{b=1}^{16}
\frac{1}{(k \hspace{-0.05cm}+\hspace{-0.05cm} p)^2 \hspace{-0.05cm}+\hspace{-0.05cm} M_{K_b}^2} \frac{1}{k^2 \hspace{-0.05cm}+\hspace{-0.05cm} M_{U_b}^2}+ \nonumber\\
&&\hspace{-1.6cm}
\sum_{b=1}^{16}
\frac{n_r}{(k \hspace{-0.05cm}+\hspace{-0.05cm} p)^2 \hspace{-0.05cm}+\hspace{-0.05cm} M_{K_b}^2} \frac{1}{k^2 \hspace{-0.05cm}+\hspace{-0.05cm} M_{S_b}^2}
-\frac{2}{(k + p)^2 \hspace{-0.05cm}+\hspace{-0.05cm} M_{K_I}^2}\frac{1}{k^2 \hspace{-0.05cm}+\hspace{-0.05cm} M_{U_I}^2 }
\nonumber\\
&&\hspace{-1.6cm}
+\frac{2}{3}\frac{1}{(k \hspace{-0.05cm}+\hspace{-0.05cm} p)^2 \hspace{-0.05cm}+\hspace{-0.05cm} M_{K_I}^2}\frac{1}{k^2 \hspace{-0.05cm}+\hspace{-0.05cm} M_{\eta I}^2}
-\frac{4}{(k \hspace{-0.05cm}+\hspace{-0.05cm} p)^2 \hspace{-0.05cm}+\hspace{-0.05cm} M_{K_I}^2}\frac{1}{k^2 \hspace{-0.05cm}+\hspace{-0.05cm} M_{S_I}^2} \nonumber
\Biggr\} .
\end{eqnarray}
Here the total contribution from pairs of the states with mass $M_U$ and $M_S$  is proportional to
\begin{eqnarray}
  (16 n_r - 4) ,
\end{eqnarray}
which vanishes when $n_r = 1/4$. The negative threshold has
nicely canceled out the unphysical threshold $K S$. The surviving thresholds
are the physical $\eta K$.

\section{Simulations and results}
\label{sec_results}
We use the MILC lattices with $2+1$ dynamical flavors of the Asqtad-improved
staggered dynamical fermions, the detailed description of the simulation parameters
can be found in Refs.~\cite{Aubin:2004wf,Bazavov:2009bb,Bernard:2010fr}.
We analyzed the $\kappa$ correlators on the $0.12$ fm MILC ensemble
of $520$ $24^3 \times 64$ gauge configurations
with bare quark masses $am_{ud}' = 0.005$
and $am_s' = 0.05$ and bare gauge coupling $10/g^2 = 6.76$
the lattice spacing $a^{-1}=1.679_{-13}^{+49}$ GeV,
which has a physical volume approximately $2.5$ fm.
The mass of the dynamical strange quark is close to its physical value
$am_s = 0.0344$~\cite{Bazavov:2009bb,Bernard:2010fr}.
The masses of the $u$ and $d$ quarks are degenerate.
In Table~\ref{tab_L0}, we list all the pseudoscalar masses used in our fits
with the exception of the masses $M_{\eta_A}$, $M_{\eta^\prime_A}$, $M_{\eta_V}$,
and $M_{\eta^\prime_V}$.
Those masses can vary with the fit parameters $\delta_A$ and $\delta_V$.

For the light $u$ quark Dirac operator $M_u$ and
the $s$ quark Dirac operator $M_s$,
we measure the point-to-point quark-line connected correlator
which is described by Eq.~(\ref{CCEQ_kappa}).
We use the conjugate gradient method (CG) to obtain
the required matrix element of the inverse fermion matrix $M_u$ and $M_s$.

In order to improve the statistics,
we place the source on all the time slices $t_s=0, \cdots, T-1$,
therefore, we perform $T=64$ inversions for each configuration
and average these correlators.
Note that the time extent of our lattices
is more than twice the spatial extent.
The rather large effort to generate propagators
allows us to evaluate the correlators with high precision,
which is important to extract the desired $\kappa$ masses reliably.

Since the $\kappa$ meson contains a strange $s$ quark and a light $u$ quark,
we should treat the $u$ quark as a valence approximation quark,
while the valence strange quark mass is
fixed to its physical value~\cite{Aubin:2004fs}.
The physical value of the strange quark mass of the lattice ensemble
used in the present work has been precisely determined
by MILC simulations~\cite{Bazavov:2009bb},
namely, $am_s=0.0344$, where $a$ is the lattice spacing.

The propagators of the $\kappa$ meson are calculated
with the same configurations using five $u$ valence quarks,
namely, we choose $am_x = 0.005$, $0.01$, $0.015$, $0.02$ and $0.025$,
where $m_x$ is the light valence $u$ quark mass.
In order to obtain the physical mass of the $\kappa$ meson, we then perform
extrapolation to the chiral (physical $\pi$ mass, obtained from PDG) limit
guided by chiral perturbation theory.
The correlators of the $\pi, K$ meson and fictitious meson $s\bar{s}$
are also measured with the same configurations
for calculating the pseudoscalar masses in Table~\ref{tab_L0}.

\begin{figure}
\includegraphics[width=8.0cm]{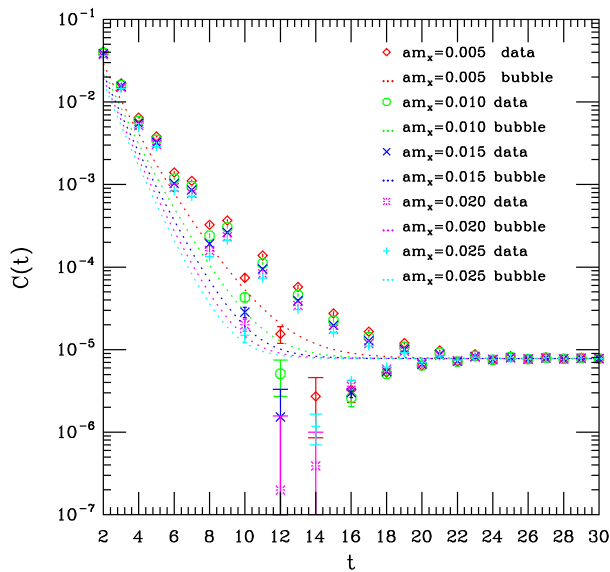}
\caption{
\label{bubble_kappa}
The $\kappa$ propagators for five valence $u$ quarks.
Overlaid on the data are their corresponding predicted bubble contributions.
}
\end{figure}

Figure~\ref{bubble_kappa} shows the $\kappa$ propagators
with five different light valence $u$ quark masses
and their predicted bubble contributions.
For our chosen MILC configurations used in the present study
we obtain the positive predicted bubble contributions,
since the small negative contribution in the bubble term is
most likely outweighed by the positive contribution
as is discussed for the $a_0$ correlators in Ref.~\cite{Prelovsek:2005rf}.
We add a constant $7.8e-6$ to all data points
and the corresponding bubble terms in y axis just for good visualization.
Figure~\ref{bubble_kappa} clearly shows that the predicted
bubble contributions dominate the $\kappa$ propagators after $t \ge 12$.

For staggered quarks the meson propagators have the generic single-particle form.
\begin{equation}
\label{sfits:ch7}
{\cal C}(t) =
\sum_i A_i e^{-m_i t} + \sum_i A_i^{'}(-1)^t e^{-m_i' t}  +(t \rightarrow N_t-t),
\end{equation}
where the oscillating terms correspond to a particle with opposite parity.
For the $\kappa$ meson correlator, we consider only one mass with each parity
in the fits of Eq.~(\ref{sfits:ch7}), namely, in our concrete calculation,
our  operator is the state with spin-taste assignment $I \otimes I$
and its oscillating term with spin-taste
assignment $\gamma_0\gamma_5 \otimes \gamma_0\gamma_5$~\cite{Ph.D:2007fzw}.
From the aforementioned discussion, we must consider the bubble contribution.
Therefore, all five $\kappa$ correlators were then fit
to the following physical model
\begin{equation}
C_{\kappa}(t) = C_{\kappa}^{\rm meson} (t)  + B_{\kappa}(t),
\label{eq:fitfcn}
\end{equation}
here
\begin{equation}
  C_{\kappa}^{ \rm meson} (t) = b_{\kappa}e^{-m_{\kappa}t}\hspace{-0.1cm} +
  b_{K_A}(-)^t e^{-M_{K_A}t} \hspace{-0.05cm}+ (t \rightarrow N_t\hspace{-0.05cm}-\hspace{-0.05cm}t), \nonumber
\end{equation}
where the $b_{K_A}$ and $b_{\kappa}$ are two overlap factors,
and the bubble term $B_{\kappa}$ in the fitting function Eq
(\ref{eq:fitfcn}) is given in momentum space by Eq.~(\ref{schpt_kappa}).
The time-Fourier transform of it yields $B_{\kappa}(t)$.

This fitting model contains the explicit $\kappa$ pole,
together with the corresponding negative-parity state $K_A$
and  the bubble contribution.
There are four fit parameters (i.e., $M_{\kappa}, M_{K_A}, b_{K_A}$,
and $b_{\kappa}$) for each $\kappa$ correlator
with a given valence $u$ quark mass $m_x$.
The bubble term $B_{\kappa}(t)$ was
parameterized by three low-energy couplings $\mu$, $\delta_A $, and $\delta_V $.
In our concrete fit, they were fixed to the values of
the previous MILC determinations~\cite{Aubin:2004fs}.
The taste multiplet masses in the bubble terms were
fixed as listed in Table~\ref{tab_L0}.
The sum over intermediate momenta was cut off when the total energy of the
two-body state exceeded $2.0/a$ or any momentum component exceeded $\pi/(4a)$.
We determined that such a cutoff gave an acceptable accuracy for $t \ge 8$.
The lightest intermediate state in bubble term is $\pi K$.
Therefore, this fitting model can remove a number of unwanted
$\pi K$ states with different tastes and slightly different energies.

\begin{figure}
\includegraphics[width=8cm,clip]{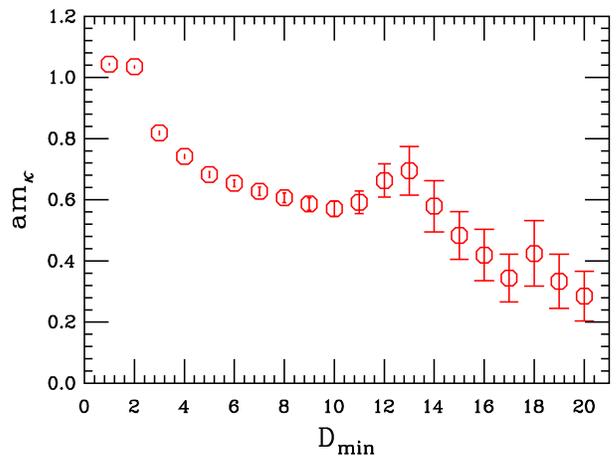}
\caption{
\label{fig:plateau}
The $\kappa$ masses as a function of the minimum time distance.
The effective mass plots will be a plateau
in time range $9\le D_{\rm min} \le 11$.
}
\end{figure}

For $am_x=0.005$, the effective mass plots of the $\kappa$ meson
are shown in Figure~\ref{fig:plateau}.
We find that the effective $\kappa$ mass suffers from large errors,
especially in larger minimum time distance regions.
To avoid possible large errors coming from the data at large minimum time
distance $D_{\rm min}$, we fit the effective mass of the $\kappa$ meson only
in the time range $9\le D_{\rm min} \le 11$,
where the effective masses are almost constant with small errors.

In our fit, five $\kappa$ propagators were
fit using a minimum time distance of $10a$.
At this distance, the contamination from the excited states is
comparable to the statistical errors,
we can neglect the systematic effect due to excited states,
therefore we can extract the mass of the $\kappa$ meson efficiently.

The fitted masses of the $\kappa$ correlators are summarized
in Table~\ref{fitted_table}.
The second block shows the masses of the $\kappa$ meson in lattice units,
and Column Four shows the time range for the chosen fit.
As a consistency check, we also list the fitted masses of
their corresponding negative parity state $K_A$ in Column Three.
We can note that the fitted values of the pseudoscalar meson $K_A$ masses
are consistent with our calculated values
in Table~\ref{tab_L0} within small errors.
Column Five shows the number of degrees of freedom (dof) for the fit.

\begin{table}
\caption{
\label{fitted_table}
The summary of the results for the fitted $\kappa$ masses.
The second block shows the $\kappa$ masses in lattice units.
The third block shows the fitted $K_A$ masses.
}
\begin{ruledtabular}
\begin{tabular}{ccccc}
$am_x$  & $am_\kappa$  & $aM_{K_A}$ & {\rm Range} & $\chi^2/{\rm dof}$  \\
\hline
$0.005$ & $0.545(26)$ & $0.3568(25)$ & $10-25$ & $12.7/12$ \\
$0.010$ & $0.590(23)$ & $0.3719(17)$ & $10-25$ & $11.8/12$ \\
$0.015$ & $0.625(21)$ & $0.3868(14)$ & $10-25$ & $11.2/12$ \\
$0.020$ & $0.653(19)$ & $0.4031(12)$ & $10-25$ & $11.0/12$ \\
$0.025$ & $0.673(18)$ & $0.4182(11)$ & $10-25$ & $11.4/12$ \\
\end{tabular}
\end{ruledtabular}
\end{table}

In order to obtain the physical mass of the $\kappa$ meson,
we carry out the chiral extrapolation of
the $\kappa$ mass $m_\kappa$ to the physical $\pi$ mass
using the popular three parameter fit with the inclusion of
the next-to-next-to-leading order (NNLO) chiral logarithms.
The general structure of the pion mass dependence of $m_\kappa$ can
be written down as
\begin{equation}
m_\kappa = c_0 + c_2 m_\pi^2 + c_3 m_\pi^3 + c_4 m_\pi^4\ln(m_\pi^2),
\end{equation}
where $c_0, c_2, c_3$ and $c_4$ are the fitting parameters,
and the fourth term is the NNLO chiral logarithms.

\begin{figure}[h!]
\includegraphics[width=8.0cm]{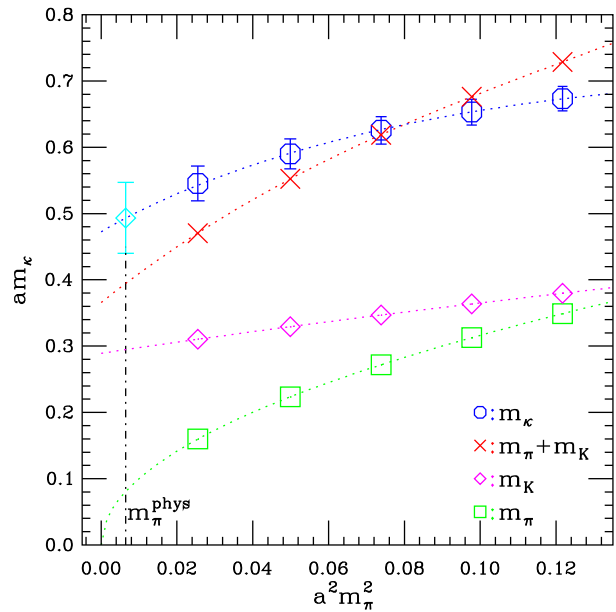}
\caption{\label{fig:kappa_limit}
Characteristics of $m_\kappa, m_K, m_\pi$ and $m_{\pi+K}$ in lattice units
as a function of the pion  mass.
The chiral limit is obtained at the physical pion mass $m_\pi$.
}
\end{figure}

We obtain the physical $\pi$ mass from PDG~\cite{Nakamura:2010zzi}
and use it as a chiral limit. In Figure~\ref{fig:kappa_limit},
we show how physical value $m_\kappa$ is extracted,
which gives $\chi^2/{\rm dof}=0.16/1$.
The blue dashed line in Figure~\ref{fig:kappa_limit}
is the linear extrapolation of the mass of the $\kappa$ meson
to the physical pion mass $m_\pi$.
The chirally extrapolated $\kappa$ mass $m_\kappa=(828\pm97)$ MeV,
which is in good accordance with the result in our previous study
on a MILC ¡°medium¡± coarse lattice ensemble~\cite{fzw:2011cpc12}.
The cyan diamond in Figure~\ref{fig:kappa_limit} indicates this value.

Using the fitting model in Ref.~\cite{Aubin:2004wf}, we extract kaon masses.
In Figure~\ref{fig:kappa_limit},
we display $m_\kappa, m_K, m_\pi$, and $m_{\pi+K}$
in lattice units as a function of the pion mass $m_\pi$.
We observe that, as the valence quark mass increases,
$\pi K$ threshold grows faster than $\kappa$ mass and,
as a consequence, $\pi K$ threshold is higher than $\kappa$ mass
for large pion mass (about $am_x \geq 0.151$),
and $\pi K$ threshold is lower than $\kappa$ meson
for small pion mass (about $am_x \leq 0.151$).
Hence, the $\kappa$ meson can decay on our lattice for small quark mass,
but for large quark mass,
the decay $\kappa \to \pi K$ is  no longer allowed kinematically,
which is in good agreement with the results in Ref.~\cite{Nebreda:2010wv}.

To understand the effects of these bubble contributions (or taste breakings),
in the present study we also fitted our measured kappa correlators 
without bubble terms.
The fitted results are tabulated in Table~\ref{fitted_table_no_bubble}.
From Table~\ref{fitted_table} and Table~\ref{fitted_table_no_bubble},
we can clearly see that the bubble terms contribute 
about $2\% - 5\%$ differences for the $\kappa$ masses.
\begin{table}[h]
\caption{
\label{fitted_table_no_bubble}
The summary of the results for the fitted $\kappa$ masses without bubble contributions.
The second block shows the $\kappa$ masses in lattice units.
The third block shows the fitted $K_A$ masses.
}
\begin{ruledtabular}
\begin{tabular}{ccccc}
$am_x$  & $am_\kappa$  & $aM_{K_A}$ & {\rm Range} & $\chi^2/{\rm dof}$  \\
\hline
$0.005$ & $0.572(19)$ & $0.3565(25)$ & $10-25$ & $12.1/12$ \\
$0.010$ & $0.615(19)$ & $0.3717(17)$ & $10-25$ & $11.0/12$ \\
$0.015$ & $0.646(19)$ & $0.3876(14)$ & $10-25$ & $10.4/12$ \\
$0.020$ & $0.669(18)$ & $0.4030(12)$ & $10-25$ & $10.2/12$ \\
$0.025$ & $0.687(18)$ & $0.4181(11)$ & $10-25$ & $10.7/12$ \\
\end{tabular}
\end{ruledtabular}
\end{table}

\section{Summary and outlook}
\label{sec_conclude}
In the present study we have extended the analyses of the scalar mesons
in Refs.~\cite{Ph.D:2006fzw,Ph.D:2007fzw},
and derived the two-pseudoscalar-meson ``bubble'' contribution to
the $\kappa$ correlator in the lowest order S$\chi$PT.
We used this physical model to fit the lattice simulation data
of the point-to-point scalar $\kappa$ correlators for
the MILC coarse ($a\approx0.12$ fm) lattice ensembles
in the presence of $2+1$ flavors of Asqtad improved staggered dynamical sea quarks,
generated by the MILC Collaboration~\cite{Aubin:2004wf,Bernard:2001av}.
We treated the light $u$ quark as a valence approximation quark,
while the strange valence $s$ quark mass is fixed to its physical value,
and chirally extrapolated the mass of the $\kappa$ meson
to the physical pion mass.
We achieved the physical mass of $\kappa$ meson with $828\pm97$ MeV,
which is very close to the recent experimental value $800\sim900$ MeV.
Probably, it may be identified with the $\kappa$ meson observed in experiments.

Most of all, we note that $\kappa$ meson is heavier than $\pi K$ threshold
for enough small $u$ quark mass.
Therefore, it can decay on our lattice for small quark mass.
This preliminary lattice simulation will stimulate people to
study the decay mode $\kappa \to \pi K$.
We are beginning lattice study of this decay channel with isospin
representation of $I=1/2$.

Since the appearance of the bubble contribution is a consequence of the fermion
determinant, an analysis of the $\kappa$ correlator in this work also provides
a direct useful test of the fourth-root recipe.
The bubble term in S$\chi$PT\ provides a useful explanation of
the lattice artifacts induced by the fourth-root approximation~\cite{Ph.D:2006fzw,Ph.D:2007fzw}.
The artifacts include the thresholds at unphysical energies and
the thresholds with negative weights.
These contributions are clearly present in the $\kappa$ channel
in our QCD simulation with the Asqtad action at $a \approx 0.12$ fm.
We find that the ``bubble'' term must be included in
a successful spectral analysis of the $\kappa$ correlator.
The rS$\chi$PT\ predicts further that these lattice artifacts
vanish in the continuum limit, leaving only physical two-body thresholds.
It would be nice to be able to investigate whether this expectation is
ruled out in lattice simulations at smaller lattice spacing.

In this work we reported our preliminary results on one lattice ensemble,
more physical one should be in the continuum limit.
We are beginning a series of numerical simulations
with the MILC fine and super-fine lattice ensembles.
Furthermore, the kappa meson is a resonance,
i.e. a state with a considerable width under strong interactions.
In order to map out ``avoided level crossings'' between the resonance
and its decay products in a finite box volume
proposed by L\"uscher~\cite{Lellouch:2001p4241},
we are beginning to measure the $\pi K$ scattering $K+\pi \to K+\pi$ channel
and the $\kappa \to K+\pi, K+\pi \to \kappa$ cross-correlators
in addition to the $\kappa \to \kappa$ correlator
~\cite{Fu:2011wc,Fu:2011xw,Fu:2011bz,Fu:2011xz}.
Hopefully, we can reliably obtain the resonance parameters 
of the $\kappa$ resonance.

\acknowledgments{
This work is supported in part by Fundamental Research Funds for the Central Universities (2010SCU23002) and by the Startup Grant from the Institute of
Nuclear Science and Technology of Sichuan University.
We would thank Carleton DeTar for kindly providing
the MILC gauge configurations used for this
work, and thank the MILC Collaboration for using
the Asqtad lattice ensemble and MILC code.
We are grateful to Hou Qing for his support.
The computations for this work were carried out
at AMAX workstation, CENTOS workstation, HP workstation
in the Radiation Physics Group of the Institute of
Nuclear Science and Technology, Sichuan University.
}

\end{document}